\documentclass[12pt,preprint]{aastex}

\newcommand{\hii}{H\,II}

\newcommand{\teff}{$T_{\!\mbox{\scriptsize \em eff}}$}
\newcommand{\teffq}{$T_{\!\mbox{\scriptsize \em eff}}^4$}

\shorttitle{A new extragalactic distance determination method}
\shortauthors{Kudritzki et al.}
\slugcomment{}

\begin{document}

\title{A new extragalactic distance determination method using
the flux-weighted gravity of late B and early A 
supergiants\footnotemark} \footnotetext[1]{Based on observations obtained
at the ESO Very Large Telescope.}

\author{Rolf P. Kudritzki, Fabio Bresolin and Norbert Przybilla}
\affil{Institute for Astronomy, 2680 Woodlawn Drive, Honolulu HI 96822}
\email{kud@ifa.hawaii.edu; bresolin@ifa.hawaii.edu; norbert@ifa.hawaii.edu}

\begin{abstract}
Stellar evolution calculations predict the flux-weighted gravity 
$g$/\teffq\/ and absolute bolometric magnitude of blue supergiants 
to be strongly correlated. We use medium resolution multi-object 
spectroscopy of late B and early A supergiants in two spiral galaxies,
NGC 300 and NGC 3621, to demonstrate the existence of such a 
relationship, which proves to be surprisingly tight. An analysis of high 
resolution spectra of blue supergiants in Local Group
galaxies confirms this detection. We discuss the application of the 
relationship for extragalactic distance determinations and conservatively
conclude that once properly calibrated it has the potential to allow for 
measurements of distance moduli out to 30.5 mag with an accuracy of 0.1 mag or 
better.
\end{abstract}


\section{Introduction}
Bright supergiants in external galaxies have been recognized as
valuable extragalactic distance indicators for a long time, since the
pioneering work of \citet{hubble36}. A number of works have attempted
to calibrate photometric and/or spectroscopic signatures of blue
supergiants (those of spectral type OBA) for this purpose, but the
uncertainty in the derived distances (typically 0.4 mag or larger in
distance modulus) have always been a major drawback for these
techniques (\citealt{humphreys88}, \citealt{tully84}).

The discovery of a wind momentum-luminosity relationship (WLR) for
blue supergiants (\citealt{kud00}) exploits the dependency of the
strength of the radiation driven winds of massive stars on stellar
luminosity, offering a potentially more accurate distance
indicator.  While hot O stars provide so far the best calibration of
the WLR (\citealt{puls96}, \citealt{puls02}), it is the visually
brightest late B and early A supergiants ($M_V\simeq-9$) which offer
the largest potential as extragalactic standard candles
(\citealt{kud98}, \citealt{kud99}). This can now be investigated for
the nearby galaxies ($D<10$ Mpc) with multiobject spectroscopy at 8-m
class telescopes, as the exploratory work of \citet{bresolin01,
bresolin02} has shown. Quantitative spectroscopy of individual BA
supergiants leads to the determination of gravities, temperatures,
metallicities and stellar wind parameters (based on the wind emission
in H$\alpha$), which are then combined to provide distances.

In this Letter we suggest a novel method, based on the absorption
strengths of the higher Balmer lines formed in the photosphere. The
concept is not entirely new, since a relationship between the
equivalent width of H$\gamma$ and $M_V$ for Galactic and Magellanic
Clouds BA supergiants, together with a temperature dependence, has
already been discussed by several authors (\citealt{petrie65},
\citealt{crampton79}, \citealt{tully84}, \citealt{hill86}).  The
recent improvements in the modelling of A supergiant atmospheres in
NLTE and the development of new diagnostic tools
(\citealt{venn95a,venn95b,venn99,venn01},
\citealt{przybilla01a,przybilla01b,przybilla02}) allows us to
determine stellar parameters with unprecedented accuracy and
reliability.  Based on these achievements, we discuss here a promising
application on a set of high-to-intermediate resolution spectra
obtained for a sample of A supergiants in the Milky Way, the
Magellanic Clouds, NGC~300 and NGC~3621, and few additional objects in
a handful of nearby galaxies.  The theoretical concept is explained in
\S 2, and we present the observational tests in \S 3 and \S 4. Future
work on supergiants in nearby galaxies is briefly summarized in \S 5.

\section{Basic concept}

Massive stars during their evolution towards the Red Supergiant stage
pass through the phase of late B and early A supergiants quickly and
with roughly constant mass and luminosity 
(\citealt{meynet00,meynet94,heger00}).
This means that in this phase the stellar gravity $g$ and effective
temperature \teff\/ are coupled through the condition
$g$/\teffq\/ $= const.$ We call $g$/\teffq\/ the ``flux-weighted gravity''.
Assuming that mass and luminosity follow the
usual relation $L \propto M^{\alpha}$, ($\alpha \sim 3$), we derive
a relationship between absolute bolometric magnitude $M_{bol}$ and the
flux-weighted gravity of the form

\begin{equation}
-M_{bol} = a\log(g/T_{\!\mbox{\scriptsize \em eff}}^4) + b
\label{fwg}
\end{equation}

\noindent
with $a$ of the order of $-3.75$. This means that for these spectral types
the fundamental stellar parameters of effective temperature and gravity
are tightly coupled to the absolute magnitude rendering the possibility
of purely spectroscopic distance determination. In the following, we refer
to Eq. 1 as the ``Flux-weighted Gravity - Luminosity Relationship (FGLR)''. 

Assuming constant luminosity and, in particular, a simple (one-exponent) power
law for the mass-luminosity relationship is, of course, a simplification. One 
might argue that the mass-loss history of supergiants  
and its dependence on stellar angular momentum and metallicity will complicate
the situation. However, we are encouraged by detailed evolutionary
calculations (\citealt{meynet00,meynet94}), which indicate for the luminosity
and mass range of late B and early A supergiants that the amount of mass lost
on the way from the main sequence is still relatively small and that
differences in mass-loss caused by stellar rotation and metallicity have no
substantial effects on the theoretical FGLRs derived.

Table 1 lists the effective temperature scale for late B and early
A supergiants based on the recent quantitative spectroscopic work cited
above. We conclude that spectral classification allows for a temperature
determination with a relative accuracy of about 4 percent. Then, for a given
spectral type (and effective temperature) the higher Balmer lines as typical
photospheric absorption lines allow for a rather precise determination of the
gravity even for only medium resolution spectra of faint 
and distant objects
(see Fig. 1). Assuming a typical relative uncertainty of 0.05 for $\log g$ we
estimate the uncertainty in $M_{bol}$ from Eq. (1) to be about 0.3 mag for a
single object. This is a very encouraging estimate. Note that at this point
we do not have to worry about absolute errors in the determination of $g$
and \teff, as long as we calibrate the method with a stellar sample at a
known distance and apply the method strictly differentially.

\section{Supergiants in NGC 300 and NGC 3621}

Recently \citet{bresolin02} studied the population of blue supergiants in
the Sculptor Group spiral galaxy NGC~300 at a distance of $\sim 2.0$ Mpc
($m-M = 26.53$, \citealt{freedman01}). Using FORS1 at the VLT medium
resolution ($\sim5\,$\AA) spectra of 70 blue supergiant candidates were
obtained and spectral types, magnitudes and colours (the latter based on the
work by \citealt{pietrzynski01}) for 62 objects were presented. 
These observations provide the ideal data for a careful test of the new method.

In our analysis we restrict ourselves to objects where we can be sure that
the higher Balmer lines (H$\gamma$, H$\delta$, etc.)  are not
contaminated by \hii~region emission, i.e. we avoid objects with clear
indication of nebular H$\alpha$ emission or with a hint of nebular emission
at H$\gamma$ above or below the 2-D stellar spectrum. In addition, we select
only spectral types within a narrow range between B8 and A4, where we know
from our recent work (\citealt{przybilla01a,przybilla01b,przybilla02}) that
our model atmosphere analysis tools are very reliable, in particular with
regard to the relative accuracy of a strictly differential study. For the
given spectral types we adopt effective temperatures according to Table~1
and determine gravities from the higher Balmer lines as displayed in Fig.~1.
We then calculate intrinsic colours with our model atmosphere code 
to determine reddening and extinction and use 
the calculated bolometric correction and the distance modulus to obtain
bolometric magnitudes.

The data set for NGC~300 is not the only one available to us. In a similar
way, \citealt{bresolin01} have used FORS1 at the VLT to study 17 objects in 
the spiral galaxy NGC~3621 at a distance of 6.7 Mpc ($m-M = 29.08$, 
\citealt{freedman01}). Applying the same selection criteria as above we can
add four more objects to the sample and apply the same spectral analysis.

The result of the test is displayed in Fig.~2, which shows a
surprisingly tight correlation, as predicted by Eq.(1). The linear
regression coefficients are $a=-3.85$ and $b=13.73$, the standard
deviation of the the residual bolometric magnitude from this
regression being $\sigma = 0.26$ mag. The objects in NGC~3621 seem to
indicate a somewhat smaller distance modulus (by 0.2 mag) than
adopted. However, we prefer to wait for forthcoming stellar photometry of both
NGC~300 and NGC~3621 with the Advanced Camera on board of HST, before
we follow up on the relative distance of these two galaxies.

\section{Objects from Local Group galaxies}

The small standard deviation obtained in Fig.~2 might be an artifact
resulting from the relatively low number of objects studied. We,
therefore, analyze additional high resolution spectra presently
available to us of a larger sample of objects in the Milky Way, the
Magellanic Clouds, M31, M33 and NGC~6822. We apply the same technique
as before, except for the three objects in the SMC, NGC~6822 and M33,
which are extremely metal poor. For those (following \citealt{venn99})
we do not rely on the spectral type, but determine \teff\/ from the
non-LTE ionization equilibrium of Mg I/II and N I/II. We note that the
inclusion of the galactic objects is somewhat problematic, as their
distances are more uncertain (\citealt{kud99,przybilla02}). We also
note that the strictly differential character of our analysis using
the same model atmosphere and non-LTE line formation codes and the
same technique in fitting Balmer line profiles is essential for the
internal accuracy of our method.  This is why we do not include the
results of other published quantitative studies of additional objects,
for which we do not have spectra at our disposal.

Fig.~3 shows the FGLR with all objects included. The regression
coefficients are very similar ($a=-3.71$, $b=13.49$) and the standard
deviation of the residual bolometric magnitude has only slightly
increased to $\sigma$ = 0.28 mag.  We note that the scatter increases
at absolute magnitudes brighter than $-8$ mag (see discussion below).

Fig.~3 also includes the relationship obtained from the evolutionary
calculations including stellar rotation and mass-loss (\citealt{meynet00}).
While the slopes practically agree (note that the stellar evolution
relationship shows a slight curvature reflecting the mild change in the
mass-luminosity exponent $\alpha$), there is a small off-set in 
log($g$/\teffq) by 0.07 dex. This can be the result of a systematic
effect in the determination of gravity or in the temperature scale, but it can
also indicate a small deficiency of the evolutionary models. In any case,
as long as an empirical calibration of the relationship with stars at known
distances is used, the accuracy of the distance determination will not be
affected.

\section{Discussion and future work}

The results presented in the previous sections are very
encouraging. The flux-weighted gravities of late B and early
A supergiants are obviously very tightly correlated with absolute
bolometric magnitude. The application of this relationship, once
properly calibrated, for extragalactic distance determinations is
straightforward. It requires multi-colour photometry of galaxies
containing a young stellar population to identify possible blue
supergiants and subsequent medium resolution ($\sim 5\,$\AA)
multi-object spectroscopy (see \citealt{bresolin01,bresolin02}) to
determine effective temperature and gravity directly from the
spectra. The spectral analysis will also yield bolometric correction
(which is small for these spectral types) and intrinsic colour so that
an accurate correction for reddening and extinction is
possible. Application of the FGLR will then provide the absolute
bolometric magnitude, which by comparison with the de-reddened visual
magnitude will give the distance modulus. Assuming a residual scatter
of $\sigma$ = 0.3 mag for the FGLR (see Fig.~2 and 3) we estimate
that with 10 supergiant stars per galaxy we can achieve an accuracy of
$0.1$ mag in distance modulus. We are confident that in one night of
observing time we can reach down to $V = 22.5$ with the existing very
efficient medium resolution multi-object spectrographs attached to
8m-class telescopes. With objects in an absolute magnitude range
between $-8$ and $-10$ mag the FGLR method appears to be applicable out
to distance moduli of $m-M = 30.5$ or even beyond.

The restriction to medium resolution spectroscopy in the blue spectral
range provides significant advantages. Most importantly, contamination from
sky and \hii~region emission is by far less critical than in the red,
which is needed as an additional spectral range, for instance, for the
WLR method, which requires the measurements of H$\alpha$ profiles with
at least $\sim2\,$\AA\/ resolution.  Moreover, the amount of observing
time for an accurate distance determination is significantly reduced,
if only spectra in the blue are required.

The accurate calibration of the FGLR (and the WLR) will become the
crucial element of future work, before the method can be applied
seriously for extragalactic distance determinations. Local Group
galaxies with well determined distances provide the ideal laboratory
for this purpose. Multi-object spectrographs with rather high spectral
resolution attached to 8 to 10m-class telescopes such as FLAMES (VLT)
or DEIMOS (Keck 2) will allow high quality spectra of large candidate
samples in each galaxy with a rather modest amount of observing time.

Such a systematic study of hundreds of blue supergiants in Local Group
galaxies will not only provide an accurate calibration of the
FGLR. It will also enable us to investigate important aspects of
stellar evolution, which are related. The most crucial ones concern
the role of metallicity and stellar rotation. Investigating stellar
evolutionary tracks at different metallicity and with different
initial rotation at the main sequence (\citealt{meynet00,meynet94}) we
find small, but noticeable effects on the theoretical FGLR. Mass-loss
in evolutionary stages prior to the blue supergiant phase depends on
metallicity (\citealt{kud00}) and has a small influence on the
FGLR. Rotation affects the strength of mass-loss and the internal
mixing processes and introduces a modification of the mass-luminosity
relationship in the blue supergiant stage. The effect of rotation
becomes larger at higher luminosities. This might be the reason for
the increase of the residual scatter at luminosities above
$M_{bol}=-8$ mentioned above.

Another very important issue is the fraction among the sample of observed blue 
supergiants evolving backwards to the blue after a previous phase as red
supergiants. Those objects are expected to have lost a
significant fraction of their mass as red supergiants and might form an
additional sequence below the observed relationship. Evolutionary calculations 
indicate that the relative number of those objects might depend crucially on
metallicity and rotation. Systematic studies of Local Group galaxies
at different metallicity, as proposed above, will allow to investigate this 
problem.

In general, the observational detection of the tight relationship
between flux-weighted gravity and absolute bolometric luminosity is a
triumph of two classical areas of astrophysics, stellar evolution and
stellar atmospheres. It confirms the general scenario of stellar
evolution with mass-loss and rotation away from the main sequence and
the predicted mass-luminosity relation. It also confirms the power and
accuracy of present day spectroscopic stellar diagnostics. It is very
satisfying to see the potential of these disciplines for a significant
contribution to quantitative extragalactic studies.


\clearpage


\clearpage

\begin{deluxetable}{lr}
\tabletypesize{\scriptsize}
\tablecolumns{2}
\tablewidth{0pt}
\tablenum{1}
\tablecaption{Adopted temperature scale\label{tscale}}
\tablehead{
\colhead{Spectral type}		& 
\colhead{\teff}}
\startdata 
B8	&	12000 \\
B9	&	10500 \\
A0	&	9500 \\
A1	&	9250 \\
A2	&	9000 \\
A3	&	8500 \\
A4	&	8350 \\
\enddata
\end{deluxetable}

\clearpage
\begin{figure*}
\plotone{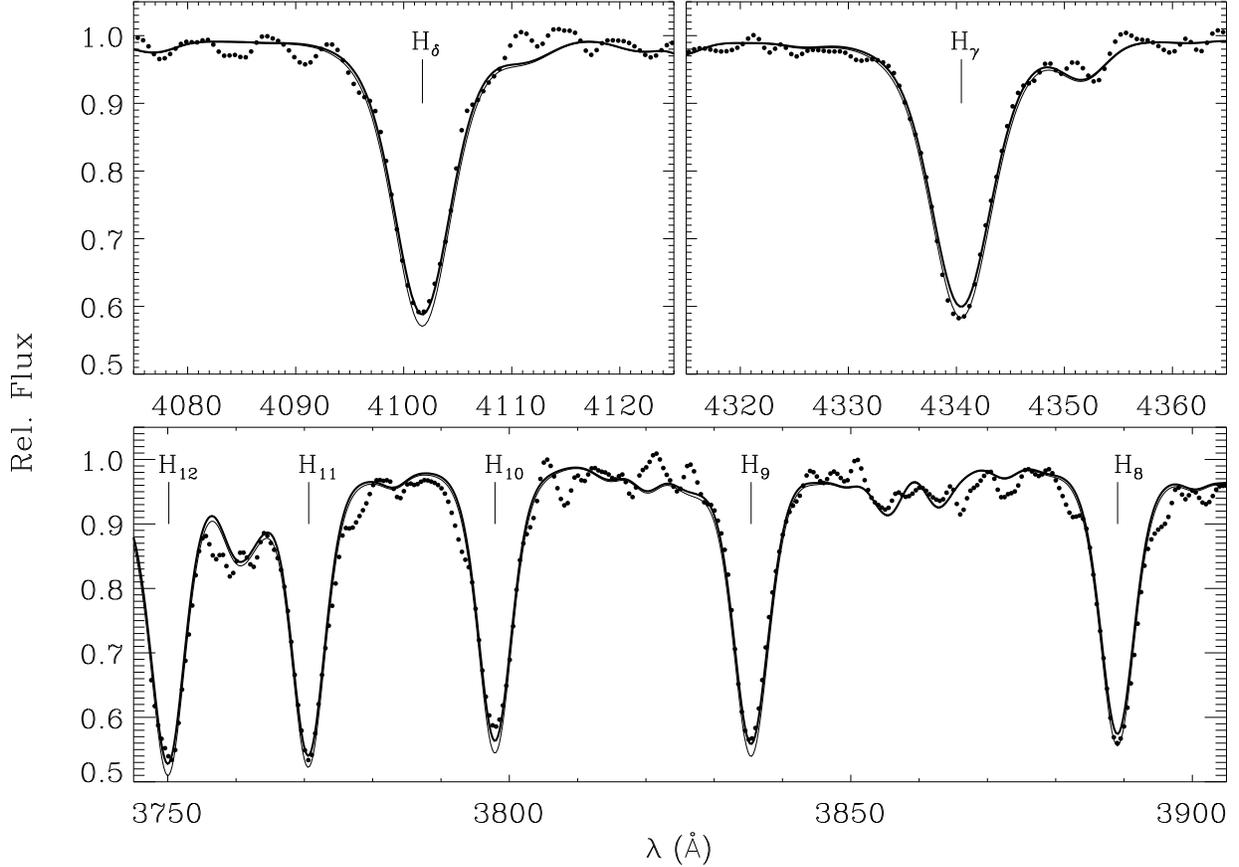}
\caption{Fit of the Balmer lines H$_{\gamma, \delta}$ and H$_{8}$ to
H$_{11}$ of the NGC 300 A0Ia supergiant C6 (see \citealt{bresolin02})
using atmospheric models with \teff\/ = 9500K and log $g$ = 1.60
(thick line) and 1.65 (thin line) (gravities given in cgs units). Note
that the use of information from many Balmer lines enhances the
accuracy of the log $g$ determination significantly. The FWHM of the
instrumental profile is $\sim 5\,$\AA, which is larger than the
intrinsic width of the Balmer lines and the line broadening through
rotation (typically $\le$ 50 km/s for A supergiants). Therefore, the
calculations are convolved with the instrumental profile. This
explains why gravity effects cannot be seen in the line wings, but
only in the cores. These effects in the line cores reflect the changes
of the integrated absorption line strength (equivalent width) as a
function of gravity.}
\end{figure*}

\begin{figure*}
\plotone{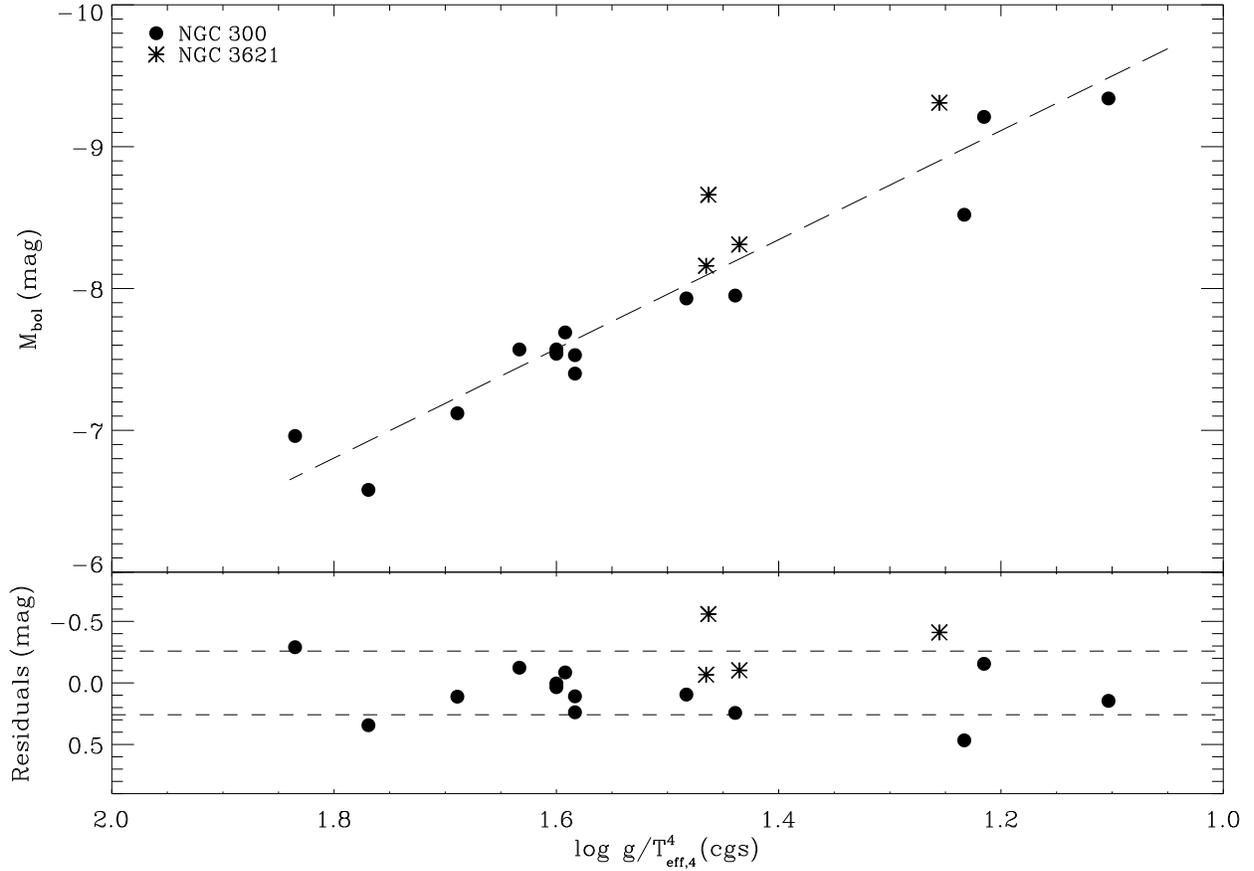}
\caption{ Absolute bolometric magnitude versus logarithm of
flux-weighted gravity of B8- to A4- supergiants in NGC 300 and NGC
3621. Note that \teff\/ is used in units of 10$^{4}$~K.  }
\end{figure*}

\begin{figure*}
\plotone{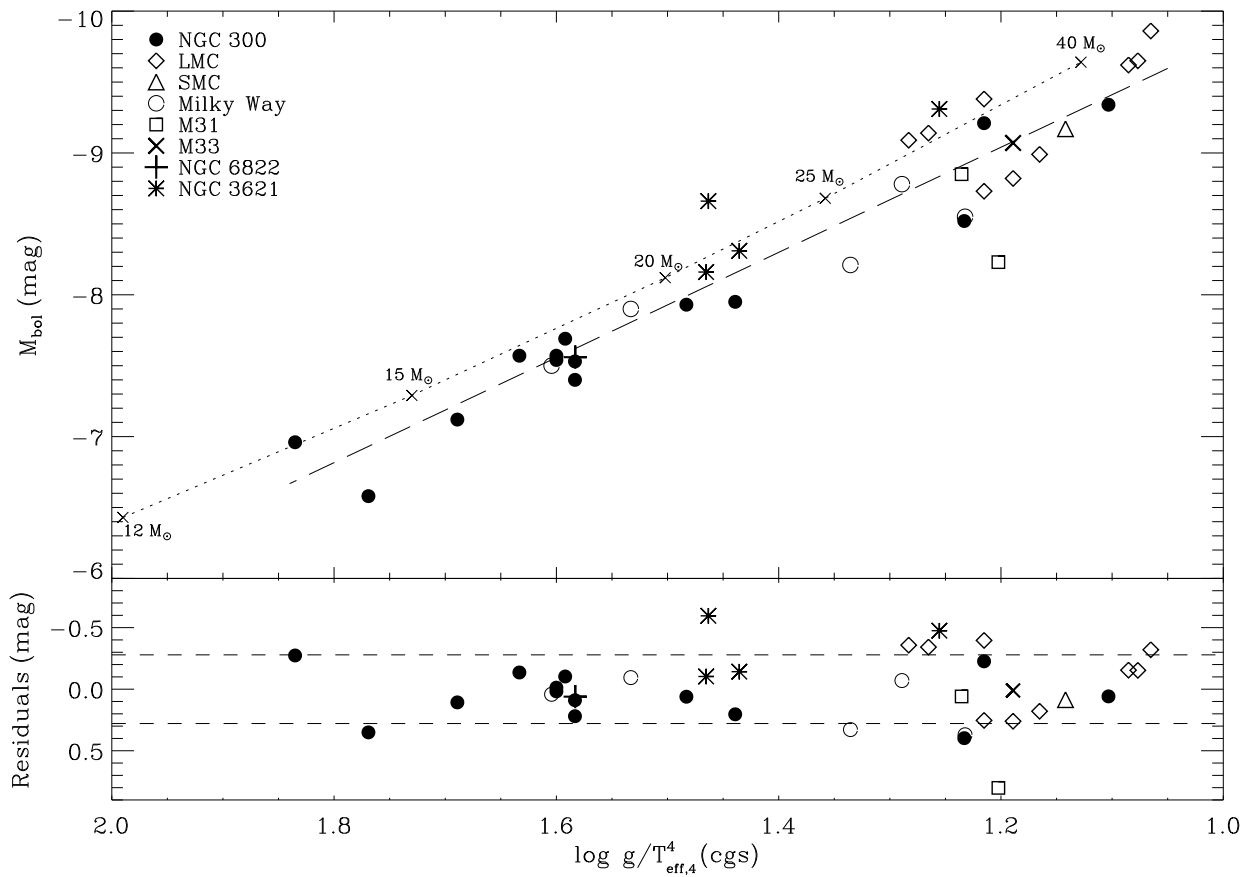}
\caption{ Same as Fig.~2 but including objects from Local Group
galaxies. The relationship obtained from stellar evolution models at
solar metallicity including the effects of rotation
(\citealt{meynet00}) is also shown and labeled with the initial
ZAMS-masses of the corresponding stellar models.  }
\end{figure*}

\end{document}